\documentclass[aps,prb,twocolumn,showpacs,superscriptaddress,amsmath,amssymb]{revtex4-1}

\usepackage{graphicx}  
\usepackage{bm}
\usepackage{xcolor}
\usepackage{gensymb}                  
\usepackage[unicode=true,pdfusetitle,
 bookmarks=true,bookmarksnumbered=false,bookmarksopen=false,
 breaklinks=false,pdfborder={0 0 1},backref=false,colorlinks=true]
 {hyperref}
\hypersetup{
 urlcolor=blue,linkcolor=blue,citecolor=blue,filecolor=blue}
 
\usepackage[T1]{fontenc}              
\usepackage{gensymb}                  

\begin{document}

\title{Structural Properties of Double-Walled Carbon Nanotubes driven by Mechanical Interlayer Coupling.}

\author{Ahmed Ghedjatti}
\affiliation{Laboratoire d'Etude des Microstructures, ONERA-CNRS, BP 72, 92322 Ch\^atillon Cedex, France}

\author{Yann Magnin}
\affiliation{Aix-Marseille University and CNRS, CINaM UMR 7325, 13288 Marseille, France}

\author{Fr\'ed\'eric Fossard}
\affiliation{Laboratoire d'Etude des Microstructures, ONERA-CNRS, BP 72, 92322 Ch\^atillon Cedex, France}

\author{Guillaume Wang}
\affiliation{Laboratoire Mat\'{e}riaux et Ph\'{e}nom\`{e}nes Quantiques, CNRS-Universit\'{e} Paris 7, 10 rue Alice Domon et L\'{e}onie Duquet, 75205 Paris Cedex 13, France}

\author{Hakim Amara}
\affiliation{Laboratoire d'Etude des Microstructures, ONERA-CNRS, BP 72, 92322 Ch\^atillon Cedex, France}

\author{Emmanuel Flahaut}
\affiliation{Centre Inter-universitaire de Recherche et d'Ing\'enierie des Mat\'eriaux (CIRIMAT), CNRS UMR 5085, Universit\'e Paul-Sabatier, 31062 Toulouse, France}

\author{Jean-S\'ebastien Lauret}
\affiliation{Laboratoire Aim\'e Cotton, CNRS, \'Ecole Normale Sup\'erieure de Cachan, Universit\'e Paris Sud, 91405 Orsay, France}

\author{Annick Loiseau}
\email{annick.loiseau@onera.fr}
\affiliation{Laboratoire d'Etude des Microstructures, ONERA-CNRS, BP 72, 92322 Ch\^atillon Cedex, France}






\begin{abstract}

Structural identification of double-walled carbon nanotubes (DWNT) is presented through a robust procedure based on the latest generation of transmission electron microscope, making possible a statistical analysis based on numerous nano-objects. This approach reveals that inner and outer tubes of DWNTs are not randomly oriented, suggesting the existence of a mechanical coupling between the two concentric walls. With the support of atomic scale modelisations, we attribute it to the presence of incommensurate domains whose structures depend on the diameters and helicities of both tubes, and where inner tubes try to achieve a local stacking orientation to reduce strain effects.

 \end{abstract}



\maketitle

\section{Introduction}

Double-walled carbon nanotubes (DWNTs) have attracted the attention of numerous scientists because their intrinsic coaxial structures give rise to exciting new applications~\cite{Shen2011, Kim2014, Moore2015a}. From a fundamental point of view, they are highly attractive since they represent the simplest system for investigating the effect of the interwall coupling on the physical properties of multi-walled carbon nanotubes. As for heterostructures built from two-dimensional crystals~\cite{Geim2013}, this interaction can be the cause of unexpected physical properties which are not well known for the moment. Therefore, a detailed understanding of the interlayer coupling is still mandatory for designing more elaborate applications of DWNTs. \\

Inter-tube electronic coupling in case of DWNTs can depend on the mutual arrangement of the tube walls defined by the inter-wall spacing and the relative rotation (or twist angle) between their hexagonal networks~\cite{Okada2003, Zolyomi2006, Moradian2007}. For instance, a strong inter-tube coupling that can induce a semiconductor-to-metal transition has been predicted for commensurate DWNTs~\cite{Zolyomi2006}. Incommensurate DWNTs were also studied~\cite{Lambin2000, Ahn2003, Uryu2005, Koshino2015}. It was recently shown that the DWNT resulting from the combination of two concentric single-walled carbon nanotubes (SWNTs) can end up with non-trivial electronic properties, depending on the twist angle~\cite{Koshino2015}. Impact of the structure of DWNT on their spectroscopic properties has also been explored. Detailed Raman studies on individual DWNTs have established that both walls are mechanically coupled via the interlayer van der Waals interaction~\cite{Dobardzic2003, Levshov2011, Liu2013, Levshov2015}. Finally, recent measurements using optical absorption spectroscopy have shown that van der Waals interaction can strongly shift optical transition energies and is highly dependent on helicity indices of each layer~\cite{Liu2014}.

In this context, an accurate knowledge of the structure of DWNTs is needed to reach a full understanding of the interactions between layers and their impact on the electronic properties. To do so, it is necessary to perform a statistical study on a large number of DWNTs to determine whether inner and outer tubes are randomly oriented each other or not, and subsequently whether they are coupled or not. Such investigations face some challenges. The first difficulty lies in the lack of synthesis routes to pure, electronically well-defined raw material~\cite{Fujisawa2016, Green2009, Moore2015b}. However the main challenge is the identification of the structure itself. A complete identification of $(n,m)$ indices of each layer of a DWNT can be extracted from the electron diffraction pattern recorded in a transmission electron microscope (TEM)~\cite{Levshov2011, Kociak2003, Hirahara2006, Deniz2010, Liu2014}. Although very powerful, it can be operated only on long, straight and isolated tubes in such a way that the electron beam illuminates solely a tube area larger than its structure periodicity. The identification of the structure can also be performed by using phase contrast high resolution imaging (so-called HRTEM technique).  This technique has suffered for a long time of a too low image resolution to provide atomically resolved images of carbon $sp^{2}$ structures. This is not longer the case with the use of TEM equipped with aberration correctors and delivering resolution below 100 $pm$. Indeed, the  direct identification of the $(n,m)$ chiral indices of SWNTs from atomically resolved images has been recently reported by Warner \textit{et al.}~\cite{Warner2011}.\\

In this article, we present a statistical study of the structure of DWNTs based on their identification from atomically resolved images recorded with a HRTEM. The DWNTs are produced by the chemical vapor deposition (CVD) technique described in~\cite{Flahaut2000} since they serve as long-standing reference samples in several works~\cite{Hertel2005, Osswald2005, DelCorro2008, Hasan2014}. We show that inner and outer tubes of DWNTs are not randomly oriented each other suggesting a strong coupling between both walls. The nature of the interwall interaction is discussed with the support of atomic scale modelisations. This leads to the conclusion that the respective orientation of the inner and outer tubes minimizes strain effects.

\section{Results and Discussion}

Our sample produced by CVD techniques~\cite{Flahaut2000, Flahaut2003} is a mixture of different structural configurations in terms of diameter and helicity. A systematic analysis of TEM images reveals that samples produced by this method contain approximately 66\% of DWNTs with a small admixture of about 20\% single-walled carbon nanotubes, and roughly 12\% triple-walled carbon nanotubes. In these experiments, the DWNTs have an outer diameter between 1.2 and 4~nm and an average inner diameter around 1.8~nm (see Figure S1 of the supporting information). Although a big care is taken to disperse properly DWNTs on TEM grids, tubes are most often entangled, so that electron diffraction can hardly be recorded from isolated tubes.

\subsection{Procedure for DWNT structure determination with HRTEM}

A complete structure identification is provided by the knowledge of the pair of Hamada indices $(n_{i}, m_{i})$@$(n_{o}, m_{o})$ where $(n_{i}, m_{i})$ and $(n_{o}, m_{o})$ stand for chiral indices of inner and outer tubes respectively. They can be extracted from geometrical parameters: the inner and outer tubes diameters, $D_{i}$ and $D_{o}$  and their respective helicities $\theta_{i}$ and $\theta_{o}$~\cite{Loiseau2006}. Here, we focus on HRTEM images~\cite{Warner2011} and examine how they can be exploited in the complex situation of a DWNT.

\begin{figure}[htbp!]
\includegraphics[width=1\linewidth]{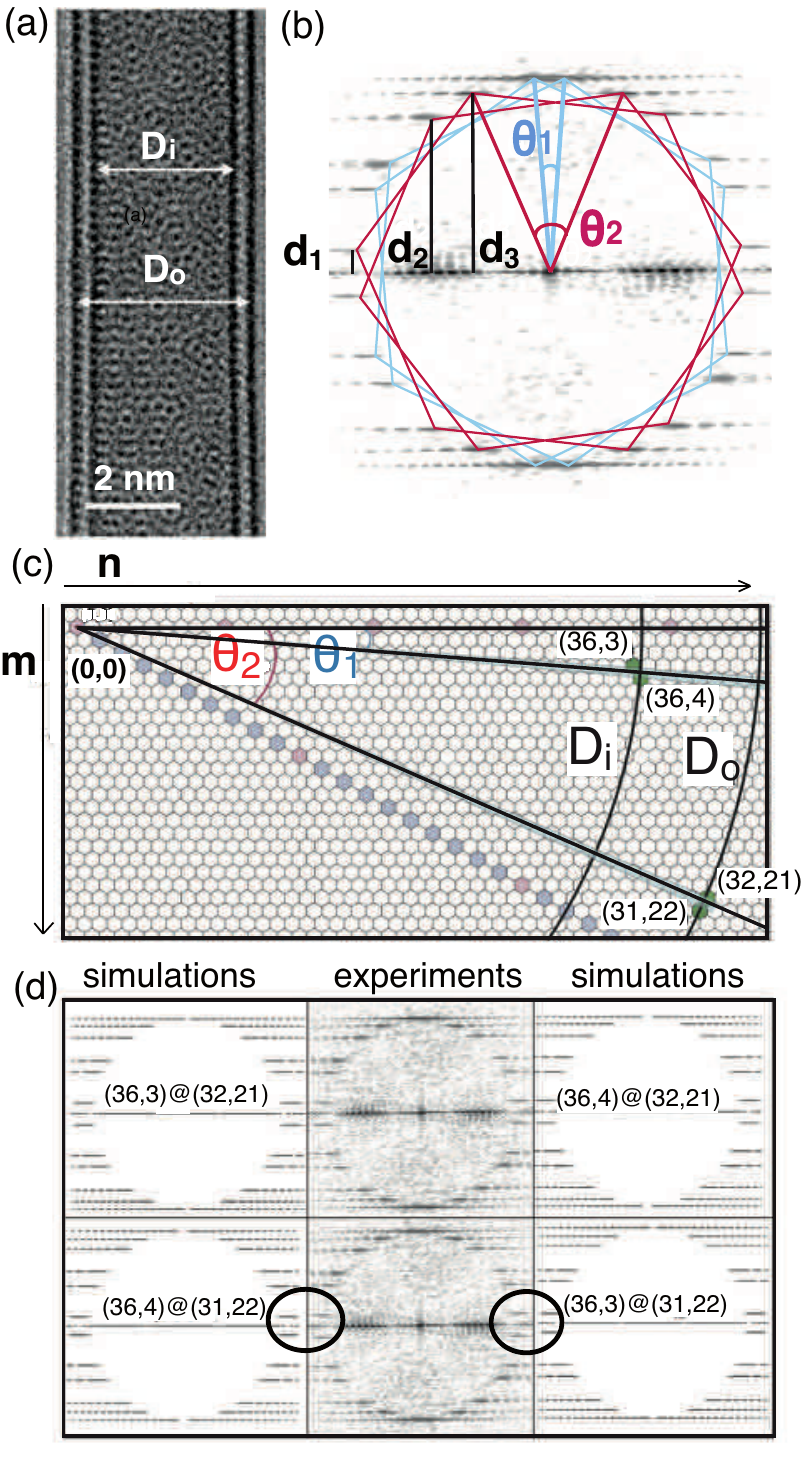}
\caption{(a) HRTEM images of a DWNT. (b) Its corresponding Fourier transform :  from the measure of the layer line spacings $d_{2}$ and $d_{3}$, chiral angles can be obtained with an error bar of $\sim0.5\degree.$ (c) Distribution of possible chiral indices after the analysis of the layer-lines. This lead to 4 configurations colored in green : (36,3)@(32,21), (36,4)@(32,21), (36,4)@(31,22) and (36,3)@(32,22). (d) Comparison of Fourier transform from HRTEM image and simulated results for previous solutions : (36,4)@(31,22) and (36,3)@(31,22) can be ruled out because some differences (marked with circles) are noticed. }
\label{Figure1}
\end{figure}

Figure~\ref{Figure1}a presents a typical atomically resolved HRTEM image of a DWNT. Basically, it displays a complex contrast Moir\'e pattern which arises from the projected view of four rotated hexagonal networks. Three rotation angles are involved: the two helicity angles ($\theta_{i}$ and $\theta_{o}$) and the twist angle between the tubes, $\Delta\theta$, defined as $||\theta_{i}| - |\theta_{o}||$.  As a result, the complexity of the Moir\'e figure hinders the direct reading of the atomic structure.

In order to achieve a fully atomic-resolved structure reconstruction from such image, we defined a data processing sequence composed of different steps. It combines analyses in real and Fourier spaces and the simulation of images based on the exact experimental TEM conditions. The first step consists in the determination of the diameters from intensity profiles related to the set of dark and bright fringes lying on each side of the tube image (see Figure S2 of the supporting information for details of the assignment)~\cite{Fleurier2009}. Diameters are determined with an error of $\sim$0.05~nm.

The second step consists in extracting helicities from the numerical fast Fourier Transform (FFT) of the image. This numerical diffraction pattern displays the same features than an experimental electron diffraction. It consists in the superimposition of two series of punctuated layer-lines related to inner and outer tubes respectively, due to discrete translation invariance along the tube axis. First order spots define four hexagons, two for the inner tube rotated each other by $\theta_{1}$ and two for the outer tube, rotated each other by $\theta_{2}$. As proposed in ~\cite{Kociak2003}, the values of the helicity angles are accurately  determined by considering spacings between the different layer lines  $d_{2}$ and $d_{3}$ as defined in Figure~\ref{Figure1}b. The chiral angle is given by:
\begin{equation}
\theta =\arctan ((2d_{2}-d_{3})/\sqrt{3}d_{3})\mbox{.} \nonumber
\end{equation}

\begin{figure}[htbp!]
\includegraphics[width=0.95\linewidth]{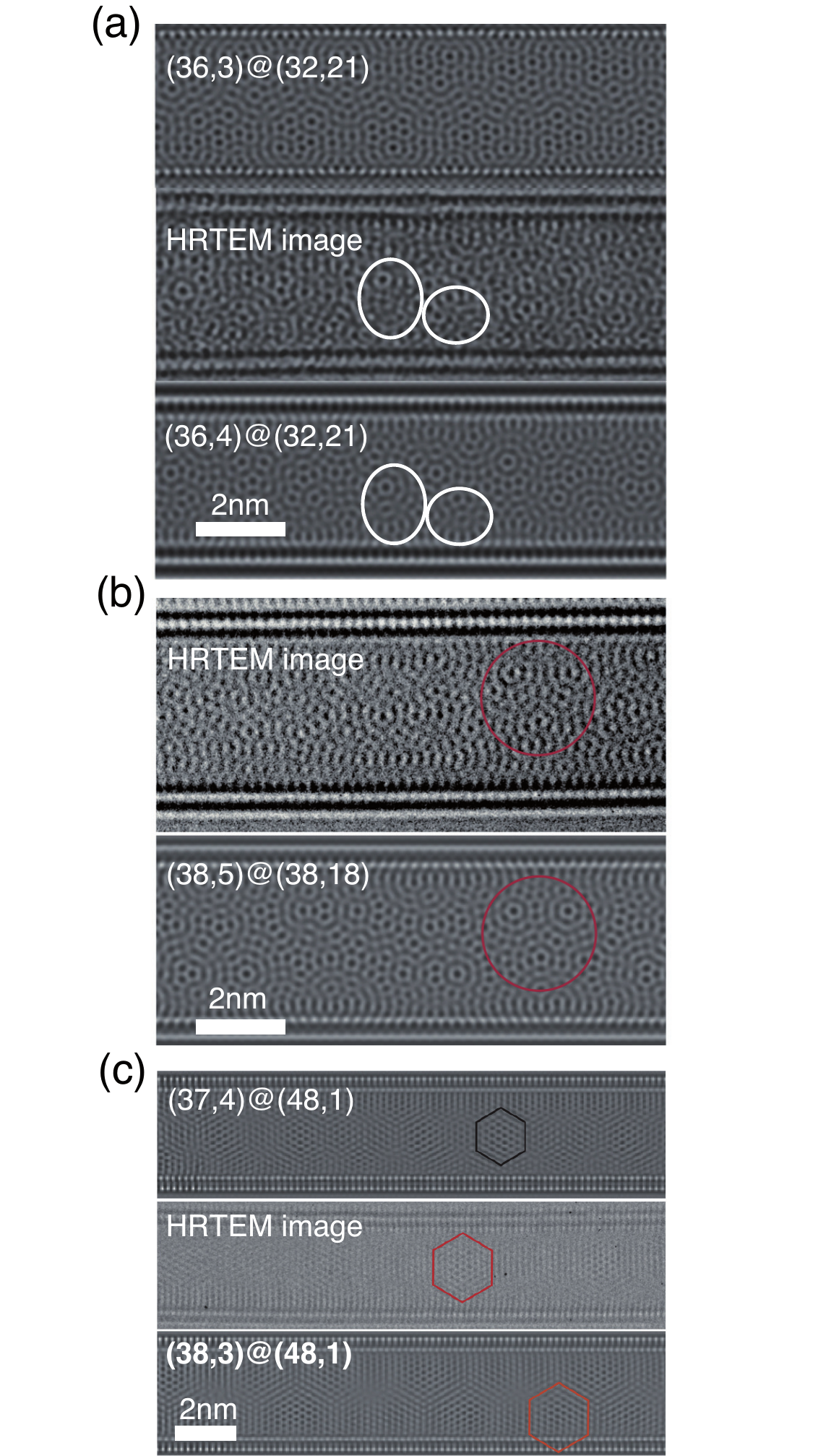}
\caption{Comparison between experimental and simulated HRTEM images where the analysis of Moir\'e patterns enables to identify the structure of the DWNT. (a) (36,3)@(32,21) and (36,4)@(32,21) DWNT ($\Delta\theta=18.0\degree$) (b) (38,5)@(3818) DWNT  ($\Delta\theta=12.24\degree$) (c) (38,3)@(48,1) DWNT  ($\Delta\theta=2.74\degree$).}
\label{Figure2}
\end{figure}

The values of the angles are obtained with an error $\sim0.5\degree$. Considering these helical angles (with error bars) and diameters of inner and outer tubes directly measured from HRTEM observations, the third step of the procedure consists in assigning all the possible $(n,m)$ tubes according to the chiral map (Figure~\ref{Figure1}c). This leads to several possible pairs of $(n_{i},m_{i})@(n_{o},m_{o})$ indices.  Then, the comparison between the experimental and the simulated FFT (see Figure~\ref{Figure1}d) allows to ruled out some configurations. The final step consists in the comparison between experimental and simulated images. The complex Moir\'e pattern is indeed very sensitive to the $(n_{i},m_{i})$ and $(n_{o},m_{o})$ couples. Indeed, changing $n$ or $m$ indices by one unit, which corresponds to a change in one helicity angle of 0.1\degree or less, can induce dramatic changes (see Figure~\ref{Figure2}a). From this procedure, only the Moir\'e pattern from the simulated HRTEM image of a (36,4)@(32,21) DWNT can fit with the experimental image in Figure~\ref{Figure2}a. Various examples corresponding to different Moir\'e patterns and twist angles are shown in Figure~\ref{Figure2}. A broad-spectrum of data is presented illustrating that any kind of configuration can be accurately determined using this procedure. Detailed examples with all the steps are shown in Figure S4 and Figure S5 of the supporting information. In this way, the chiral indices of DWNTs can be determined unambiguously from our robust procedure. \\

\subsection{Statistical analysis of DWNT helicities}

In order to emphasize a possible correlation between the DWNT layers, we examine whether there are preferred combinations of inner and outer tubes by analyzing the statistical distributions of different structural parameters of $\sim$70 isolated DWNTs. Let us first consider the interlayer distance $\Delta r$, where $\Delta r=(D_{o}-D_{i})/2$. As seen in Figure S6, values of $\Delta r$ are distributed over a relatively wide range of 0.30~nm to 0.40~nm close to that of bulk graphite ($\sim$0.34 nm). More interestingly, the results show no significant correlation between $\Delta r$ and $D_{o}$ in agreement with previous works~\cite{Hashimoto2005, Hirahara2006} (see Figure S6 of the supporting information).
\begin{figure}[htbp!]
\includegraphics[width=1\linewidth]{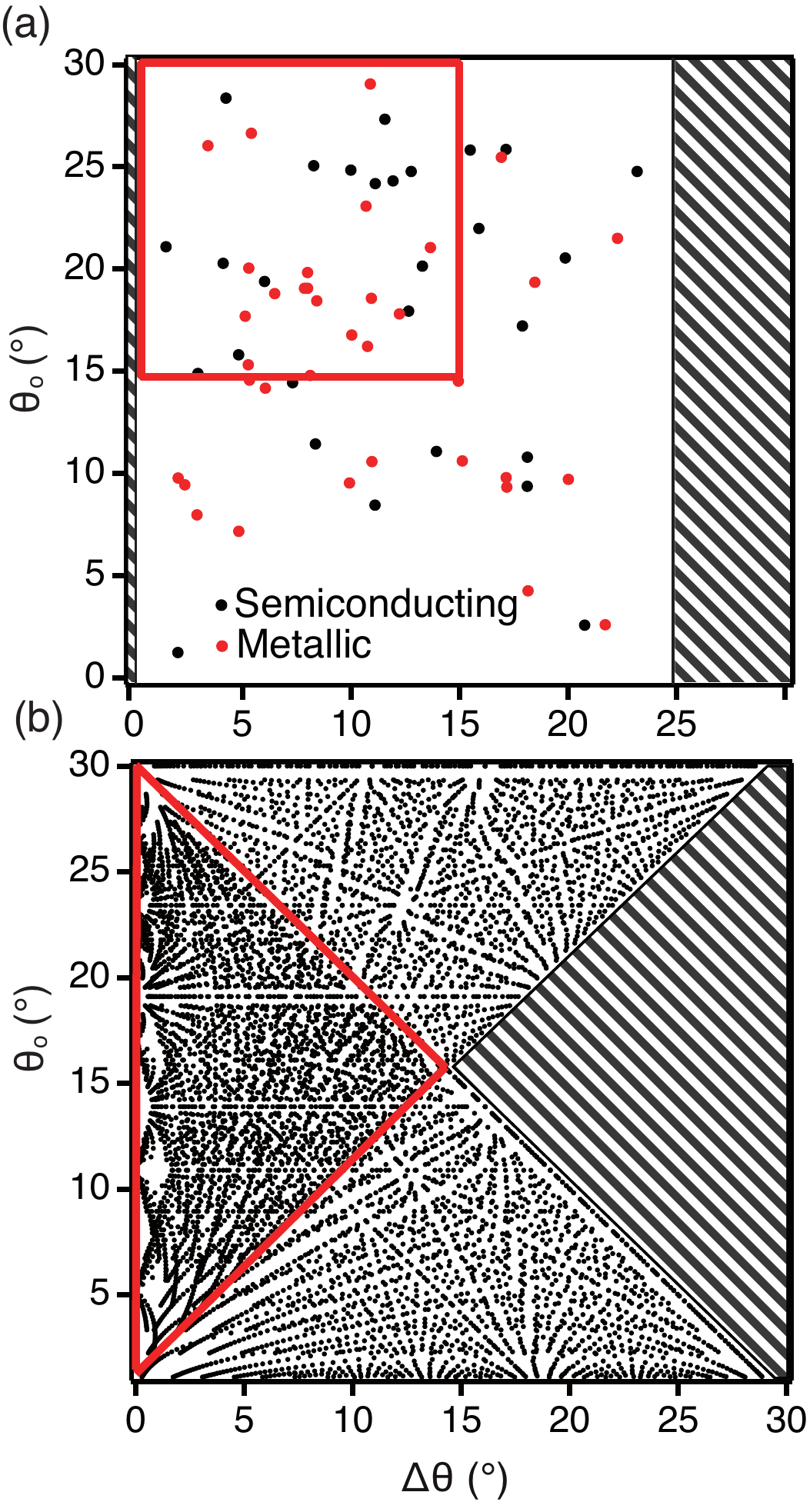}
\caption{(a) Statistical analysis of $\sim$70 DWNT helicities following our TEM procedure. (b) Random distribution of $\Delta\theta$ for all $(n_{i},m_{i})$ and $(n_{o},m_{o})$ in the ranges $1.5 <D_{m}< 4.0$ nm and 0.30<$\Delta r$<0.40 nm. In all cases, favored (red delimited area) and non observed (dashed grey area) configurations are marked.}
\label{Figure3}
\end{figure}

Then, the apparent differences in the chiral angles between inner and outer tubes are examined by analysing the relationship between the helicities of outer tubes $\theta_{o}$ and $\Delta\theta$.  The chirality distribution was discussed with respect to $\Delta\theta$ which implies the absolute value of the chiral angle, since we could not adequately distinguish right- or left-handed chirality from the experiment. As visualized in Figure~\ref{Figure3}a, the distribution of the chiral indices of the DWNTs is not homogeneous. Indeed, configurations corresponding to areas filled in grey are not observed, i.e., $\Delta\theta = 0\degree$ and $\Delta\theta > 25\degree$. The first exclusion zone is not so surprising since commensurate DWNTs ($\Delta\theta = 0\degree$) are rarely (or never) observed experimentally because it is unlikely to have two commensurate SWNTs with the appropriate radius difference for the formation of a DWNT~\cite{Liu2014, Kociak2003, Kociak2002}. As for the second one, it suggests that inner and outer tubes are strongly correlated in such a way to avoid  twinning angles where $\Delta\theta > 25\degree$. Besides, some configurations are particularly favored and are highlighted by a red square in Figure~\ref{Figure3}a where $\sim50\%$ of the nanotubes are observed. This zone fulfills two conditions: both helicities are near armchair and $\Delta\theta < 15\degree$. In order to demonstrate the particular features of the relationship between $\theta_{o}$ and $\Delta\theta$, we now consider the random orientation of both layers as reference data (see Figure~\ref{Figure3}b). The distribution is calculated for all $(n_{i}, m_{i})$ @ $(n_{o}, m_{o})$ in the ranges of $1.5<D_{o}<4$~nm and $0.30<\Delta_{r}<0.40$~nm. As seen in Figure~\ref{Figure3}b, favored and non observed configurations are expected to be distributed in red and grey triangles respectively. Moreover, our calculations show an homogeneous and symmetric repartition with respect to 15\degree~for both axes, in strong contrast with our experimental findings. This supports the conclusion that inner and outer tubes are not randomly oriented each other. It is worth mentioning that the correlation relates only to $\Delta\theta$ and has no impact on the interlayer spacing as shown in Figure S6 of the supporting information. 

Using our criterion, we have also analyzed data found in the literature where DWNTs were prepared using arc discharge method~\cite{Hirahara2006} and an other CVD technique~\cite{Liu2014} to confirm the intrinsic character of our results. In these works, structural analysis have been performed by using electron diffraction. One can note that the relationship between the helicities and $\Delta\theta$ has not been investigated in these previous works making impossible the conclusions discussed here. By analysing those experimental results with our procedure, similar conclusions to ours can be proposed in terms of favored (red square) and forbidden configurations (grey area)(see Figure S7). As a result, it can be concluded that the orientations of the hexagonal carbon network between the inner and outer tubes of DWNT are not independent, and that this results does not depend neither on the synthesis technique nor on the method of analysis. \\

\subsection{Mechanical coupling between layers}

\begin{figure}[htbp!]
\includegraphics[width=0.85\linewidth]{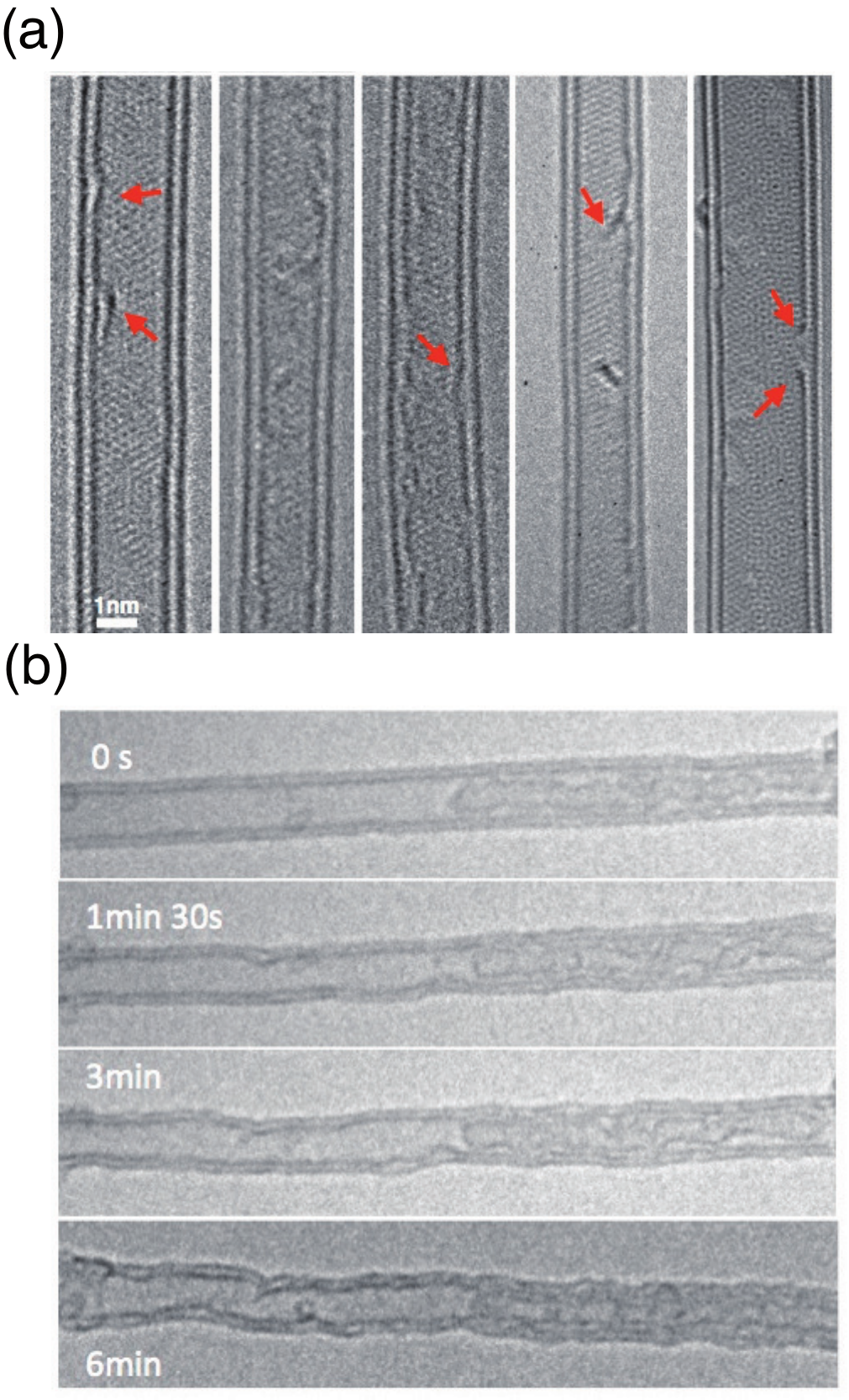}
\caption{(a) HRTEM image of defected inner tubes in some isolated DWNTs. Local deformations of inner tubes are highlighted by red arrows. (b) Structural changes of DWNTs under electron irradiation.}
\label{Figure4}
\end{figure}

The next step is to identify the nature of the coupling, i.e., electronic and/or mechanical which is responsible to previous observations. From the electronic point of view, all the SWNT that form the DWNT can be classified according to their electronic nature by considering the chiral indices $(n_{i},m_{i})$ and $(n_{o},m_{o})$. However, due to inter-layer coupling, electronic properties of DWNTs can differ from those of the constituent SWNTs. By using theoretical arguments developed in~\cite{Koshino2015} that take into account this coupling, the electronic properties of the nanotubes characterized above have been determined. As seen in Figure~\ref{Figure3}a, the distribution in terms of electronic population is relatively homogeneous. Therefore, it can be concluded that the inter-wall coupling giving rise to favored and forbidden configurations is unlikely to be driven by electronic effects. \\

Besides, in some peculiar isolated DWNTs, anomalous structures have been occasionally found suggesting that a mechanical coupling between walls can exist. Indeed, different HRTEM observations have revealed the presence of defected DWNTs where only the inner tubes are damaged. As seen in Figure~\ref{Figure4}a, structural changes of the confined tube characterized by local deformations (highlighted by red arrows) are observed. In the present cases, the interlayer distance shows a variation along the tube axis. Indeed, the inner tube bends locally to decrease the distance with the outer wall. Previous works have already revealed the deformation of neighbouring two DWNTs due to van der Waals interactions present in bundle~\cite{Hashimoto2005}. In such situation, both walls sustained damage. This is clearly not the case here since only isolated DWNTs are considered. Moreover, electron beam produced by TEM could also be at the root of these structural modifications. To test this assumption, TEM observations have been performed with a Philips CM20 operating at 120 kV. Figure~\ref{Figure4}b shows structural changes of a DWNT during the electron irradiation. Both walls are clearly damaged, indicating that when the deformation is observed solely on the inner tube, it is not due to the electron irradiation of the microscope. As a result, the presence of defected inner tubes suggests that a mechanical coupling exist between walls. A mechanical coupling has already been reported using Raman spectroscopy~\cite{Levshov2011, Levshov2015} and for the first time, is directly evidenced in the case of a DWNT by TEM investigations. \\

\subsection{Atomic scale simulations}

We have performed atomic scale modelisation in order to get insight into the coupling between walls depending on whether the constituent nanotubes are commensurate or not. In the layered graphitic sheets, the interlayer interaction is dominated by the long-ranged van der Waals interaction. Therefore, empirical methods capable of predicting the equilibrium  distance at the  van der Waals  distances  are  needed  for studying large graphitic systems. In the present simulations, the atomic interaction is described by a potential developed by Che \textit{et al}~\cite{Che1999}.\\

\begin{figure}[htbp!]
\includegraphics[width=1.00\linewidth]{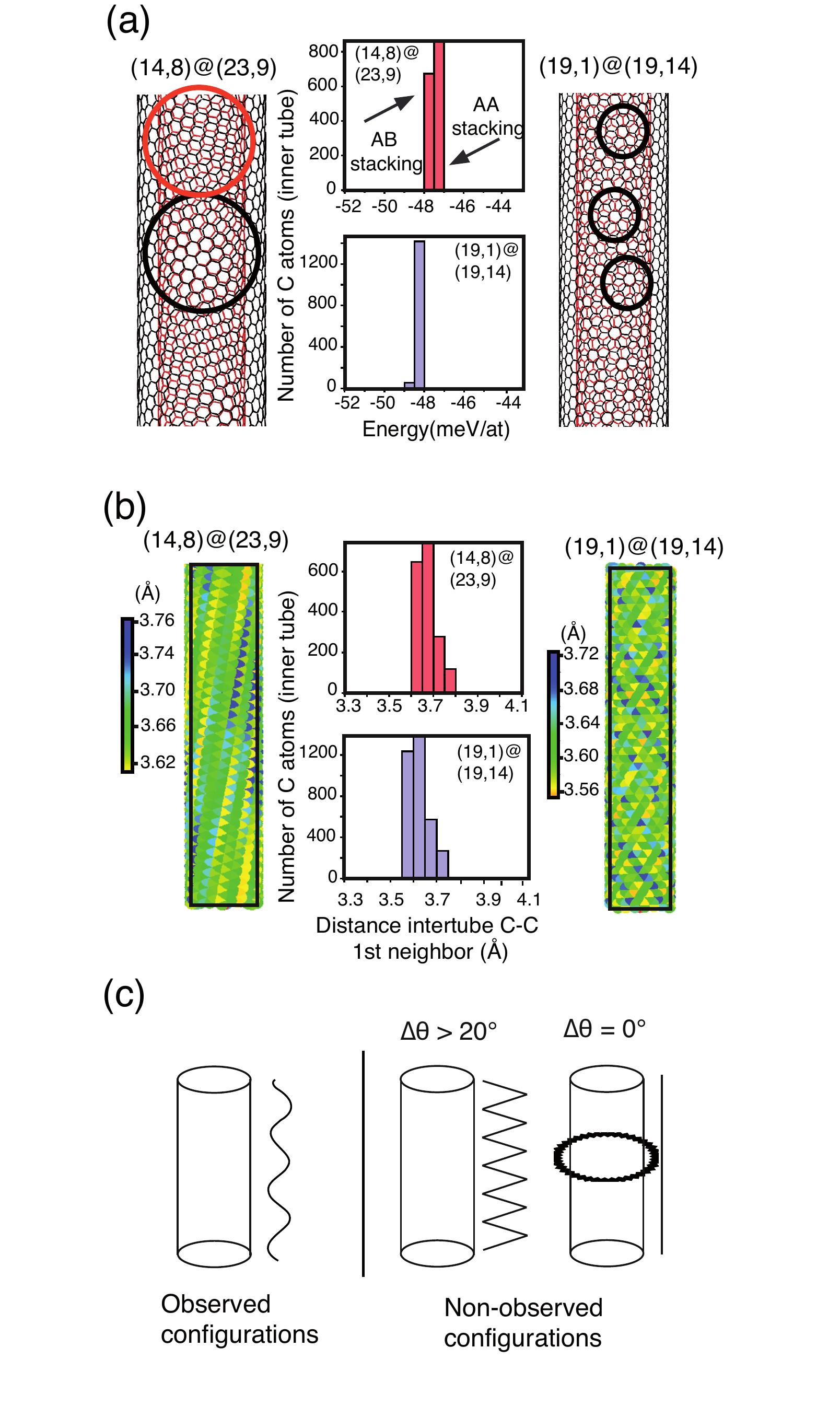}
\caption{(a) Analysis of local energies of a (14,8)@(23,9) DWNT ($\Delta\theta=5.23\degree$) and (19,1)@(19,14) DWNT ($\Delta\theta=22.46\degree$) in form of histogram plots (middle). Different stackings (AA or AB) are highlighted by circles (left and right). (b) Analysis of the C-C first neighbors intertube distances in form of histogram plots (middle) and spatial distribution along the tube  (left and right). (c) Sketchs to illustrate the different types of roughness between walls. }
\label{Figure5}
\end{figure}
First, relaxed structures have been obtained after performing rigid relaxation including translations and rotations of the inner tube while the outer one is kept fixed. Using this approach, commensurate as well as incommensurate DWNTs containing a large number of atoms ($\sim$1000 to 10 000 atoms) are studied. Then, we analyse in an histogram form the local energies of carbon atoms of the inner tube to determine which C atoms gain some energy. As an example, results for the (14,8)@(23,9) DWNT ($\Delta\theta=5.23\degree$), which has been observed experimentally, are plotted on Figure~\ref{Figure5}a. Two populations of carbon atoms are identified. They correspond to a Moir\'e pattern displaying two kinds of local stacking. On one hand, regions where neighbors atoms are almost on the top of each other, corresponding to the so-called AA stacking, are characterized by the highest interaction energy. On the other hand, since the walls are rotated, the well-known Bernal AB stacking is also observed, giving rise to the lowest interaction energy. These results are in agreement with \textit{ab initio} calculations showing that AB-stacked bilayer graphene is the most stable structure~\cite{Mostaani2015, Bucko2016}. Considering now the case of unexpected (19,1)@(19,14) DWNT ($\Delta\theta=22.46\degree$), only one population of C atoms is evidenced (AB stacking), where their energies are slightly different with the previous example since diameters are not the same (see Figure~\ref{Figure5}a). These configurations show a more uniform pattern where AA stacking is no more present. Other examples displaying the same local energy distributions are discussed in the Supp. Materials. Despite these examples, local energy differences between stackings are not significant enough (less than 1 meV/at) to explain the favored and forbidden configurations highlight by our experiments.

Consequently, we have then considered geometrical arguments by investigating the spatial distributions of C-C first neighbours intertube distances. This can be seen as a signature of the roughness between walls and therefore can be very helpful to identify a possible mechanical coupling in DWNTs.  As seen in histogram plots presented in Figure~\ref{Figure5}b, both configurations display the same C-C first neighbours intertube distance distribution. However, their spatial distributions along the inner tube are strongly different. In the case of the observed (14,8)@(23,9) DWNT multiple domains of two different Bernal-stacked configurations (AB vs AA stacking) coexist. When looking at the C-C distances mapping, this DWNT presents a smooth variation (illustrated by a wavy profile in Figure~\ref{Figure5}c) pointing out that roughness between both walls is soft. This behaviour has also been identified in case of an observed (11,10)@(20,11) DWNT as shown in  Figure~S8. Both areas are present and the roughness between walls is relatively poor. We thus show that the inner tube is subject to weak stress effects. We now focus on the unobserved (19,1)@(19,14) DWNT. As seen in Figure~\ref{Figure5}b, the mapping along the inner tube differs strongly from the one of the (14,8)@(23,9) DWNT. Indeed, our analysis shows an irregular pattern highly pronounced where spatial distributions are sharp and discrete (see sketch in Figure~\ref{Figure5}c). This indicates that inner and outer tubes interact strongly resulting in rough intertube spacings. We can therefore attribute the non observation of the (19,1)@(19,14) DWNT to the strain effects on the inner tube which tend to prevent this kind of stacking. Similarly, such mechanical coupling can explain the lack of commensurate DWNTs for a $\Delta\theta$ equal (or close) to 0$\degree$ (see Figure~S8 of the supporting information). \\

\begin{figure}[htbp!]
\includegraphics[width=1.0\linewidth]{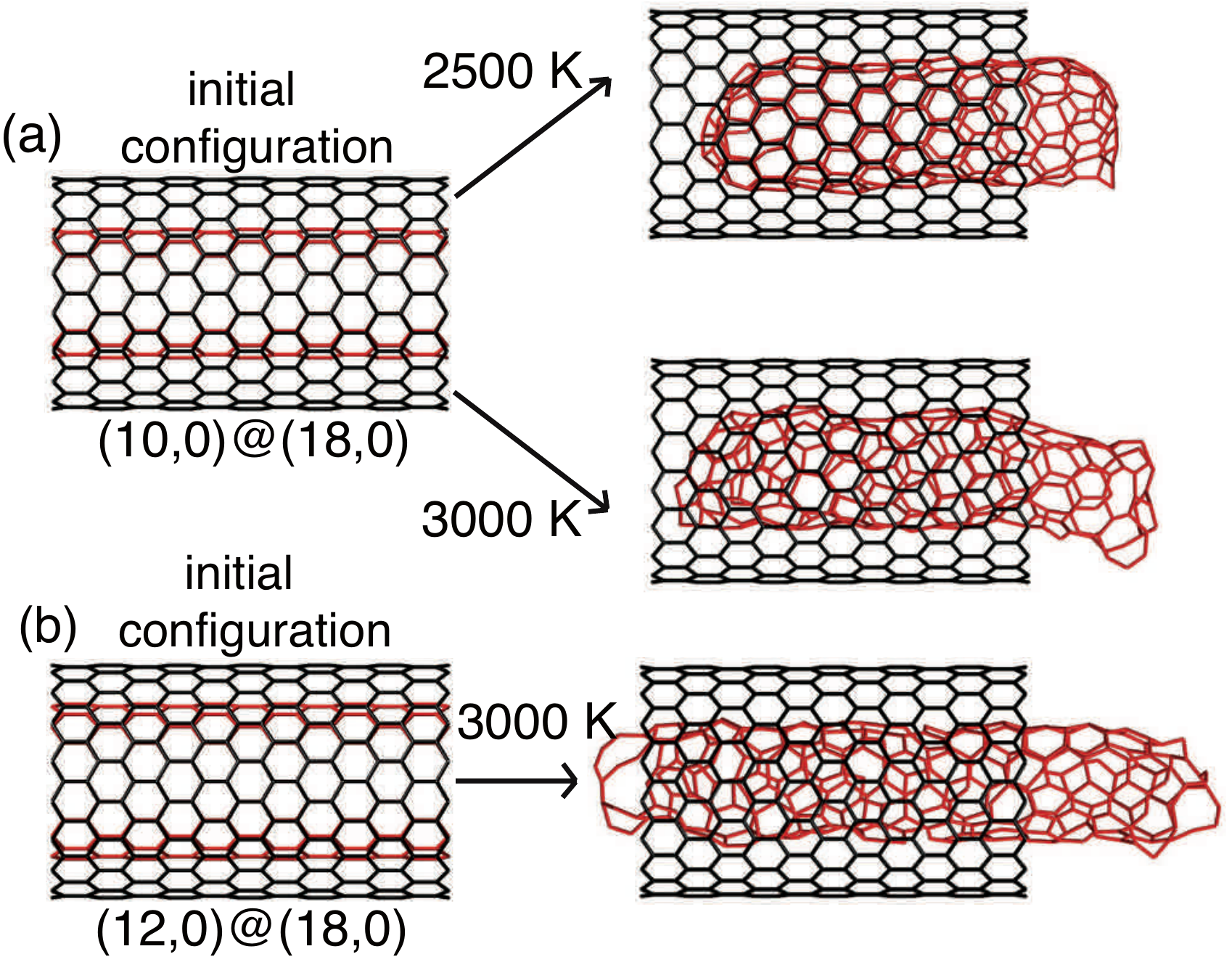}
\caption{(a) Relaxed configurations of (10,0)@(18,0) DWNT after MC simulations at 2500 K and 3000 K (b) Relaxed configuration of (12,0)@(18,0) DWNT after MC simulation at 3000 K}
\label{Figure6}
\end{figure}

Lastly, we introduce the out-of-plane degrees of freedom by using Monte Carlo (MC) simulations~\cite{Frenkel2002} to relax the structures and analyze the non stability of specific DWNTs. The idea is to start from different cases not found experimentally and submit them to high temperatures up to 3000 K, which is a typical temperature used in graphitization processes~\cite{Setton2002}. The system is then able to overcome high energy barriers and reach new states corresponding to equilibrium configurations. Using this procedure, the mechanisms explaining the stability of different DWNTs can be studied in detail.

We present two representative DWNTs: (10,0)@(18,0) and (12,0)@(18,0) which are both commensurate with $\Delta\theta=0\degree$ and not supposed to be stable according to our HRTEM observations. We have deliberately chosen tubes with very small interlayer distances $\Delta r$ (0.31~nm and 0.27~nm, respectively) to emphasize interlayer coupling and obtain relaxed structures in a reasonable, although long, CPU time. To mimic experimental observations where only the inner tube is subjected to structural modifications, outer shells are kept fixed and no periodic boundary conditions have been applied along the tube axis in order to allow the inner tube to relax completely.

Figure~\ref{Figure6}a shows the final states of the MC runs for a (10,0)@(18,0) DWNT at two temperatures ($T=2500$~K and $T=3000$~K) containing 672~atoms. At $2500$~K, the spontaneous closing of the inner tube into a graphitic like dome can be observed. This is not surprising since these edge relaxations are due to the presence of unstable dangling bonds. More interestingly, the inner tube moves along the tube axis (displacement $\sim$0.2-0.3 nm) to minimize its interaction with the outer shell, and the local stacking changes continuously during the simulation. This shows the non stability of the initial structure since starting from a AA stacking the final configuration corresponds to a AB one. At higher temperature ($T=3000$ K), spectacular structural changes are noticed. The diameter and chiral angle are strongly modified as seen in Figure~\ref{Figure6}a showing that configurations with $\Delta\theta=0\degree$ are not stable at all. It is reasonable to think that the equilibrium configuration should correspond to DWNT with structural parameters corresponding to favored configurations as discussed in Figure~\ref{Figure3}. However, the relaxed structure presents important distortions and a lot of defects preventing a complete determination of its chirality. Despite that, our simulations emphasize that strain effects on the inner tube exist leading to forbidden structures, as observed experimentally. In the case of (12,0)@(18,0) DWNT, same conclusions can be drawn with stronger evidences due to the small interlayer distance which increases the effects (see Figure~\ref{Figure6}b). As an example, diameter of the inner tube varies from 9.5~\AA \ to 6.5~\AA. To conclude, an important issue derived from these results is that they confirm that the inner tube can be subjected to stress effects due to a interlayer coupling explaining the non stability of particular configurations and the deformation of inner tubes.

\section{Conclusion}

In summary, we have shown that both layers constituting DWNTs are not randomly oriented. Their structural properties are mainly driven by a mechanical interlayer coupling where the inner tube tries to achieve a local stacking in order to reduce strain effects. This can be achieved during their synthesis since it has been suggested that MWNTs grow by a layer-by-layer mechanism~\cite{Iijima1993, Guo1995, Rodriguez-Manzo2007}. In this scenario, the outer tube mights start to grow; then inner walls are formed which can be stabilized by a lip-lip interaction~\cite{Charlier1997}. By repeating this growth process, a nanotube can grow in length as well as in thickness. Recently some \textit{in-situ} TEM observations demonstrated this sequential nucleation processes in case of DWNTs by using crystalline Pt as catalyst particle~\cite{Zhang2016}. Our work is in complete agreement with this proposed mechanism and highlights the directional correlation between two adjacent graphitic layers in a DWNT when it grows, to accommodate at best strain effects on the inner tube. Meanwhile, the presence of favored configurations shed new light on the structural control of DWNT. Indeed, the main hurdle in the development of a DWNT-based technology is to control their structure and more precisely their diameter and chirality. For this reason, several groups are working on the subsequent processing and sorting of raw material~\cite{Green2009, Moore2015b}. By focusing on favored structures, realizing the ultimate goal of controlling chirality during sorting will be facilitated enabling their use for a wide variety of potential applications.


\acknowledgments

The research leading to these results has received funding from C'Nano Ile-de-France (project Biptec). Y. M and H. A. thank the European Union Seventh Framework Programme (FP7/2007-2013) under grant agreement n° 604472 (IRENA project).  J. S. L. were
partly funded by Institut Universitaire de France. G. W acknowledges the support of the French Agence Nationale de la Recherche (ANR) under contract reference ANR-11-BS10-009.  Authors wish to thank the METSA research foundation for giving access to the Cs-corrected TEM of MPQ-Paris Diderot laboratory. We are also grateful to the Region Ile-de-France for convention SESAME E1845 for the support of the JEOL ARM 200F electron microscope installed at the Paris Diderot University. Authors acknowledge Drs M. Kociak (DiffractX) and Y. Le Bouar for using their softwares. Drs F. Ducastelle, L. Marty and P. Poncharal are acknowledged for helpful discussions.



\newpage

\begin{widetext}

\section*{Supplementary Material of : Structural Properties of Double-Walled Carbon Nanotubes driven by Mechanical Interlayer Coupling.}

\renewcommand\thefigure{S\arabic{figure}}
\setcounter{figure}{0}

\textbf{Methods}\\
Double-wall carbon nanotubes were grown using CVD method  based on a thermal decomposition of CH$_{4}$ on Co:Mo-MgO~\cite{Flahaut2000, Flahaut2003}. TEM observations have been performed with an aberration-corrected microscope, the JEOL-ARM-200F with a spatial resolution of 80 $pm$ for an accelerating voltage of 80 kV. Simulated TEM images have been calculated within dynamical theory with a commercial code (JEMS~\cite{Stadelmann1987}) and a homemade software (DiffractX~\cite{Kociak2003, Lambin1997}).We have used the multislice approach~\cite{Kirkland2009} based on scattering factors given by Peng \textit{et al.}~\cite{Peng1996} and aberration coefficients corresponding to the JEM-ARM-200F~\cite{Ricolleau2012}. \\

The cohesion in layered graphitic structures is a combination of long-ranged van der Waals and short-ranged orbital overlap contributions. The first term implies a cutoff distance around 20 \AA~to reproduce correctly the interlayer energy in graphite~\cite{Che1999}. To limit the CPU time of our MC simulations, only commensurate tubes have been considered when complete atomic relaxations have been investigated. The convergence of the total energy as a function of Monte Carlo steps is controlled and the simulation is stopped when the total energy no longer varies on the average, which implies that the system has reached a Gibbs energy minimum. Typical runs consist in $10^{3}$ external Monte Carlo loops, each of them randomly performing $10^{3}$ atomic displacements trials. The different tubes studied here were subjected to temperatures ranging from 1000 to 4500 K. At low temperature ($T<2500$ K) no obvious structural modification are observed since the system is trapped in a local minima. On the other hand, close to vaporization conditions ($T> 4000$ K), the tube is completely destroyed. To avoid this difficulty, simulations are performed at different temperatures ranging from 2500 K to 3500 K.

\newpage

\textbf{The DWNT samples under study}\\
In Figure \ref{figS1} we report the TEM analysis of our samples in term of population and diameter of tubes. 

\begin{figure}[htbp!]
\begin{center}
\includegraphics[width=0.7\linewidth]{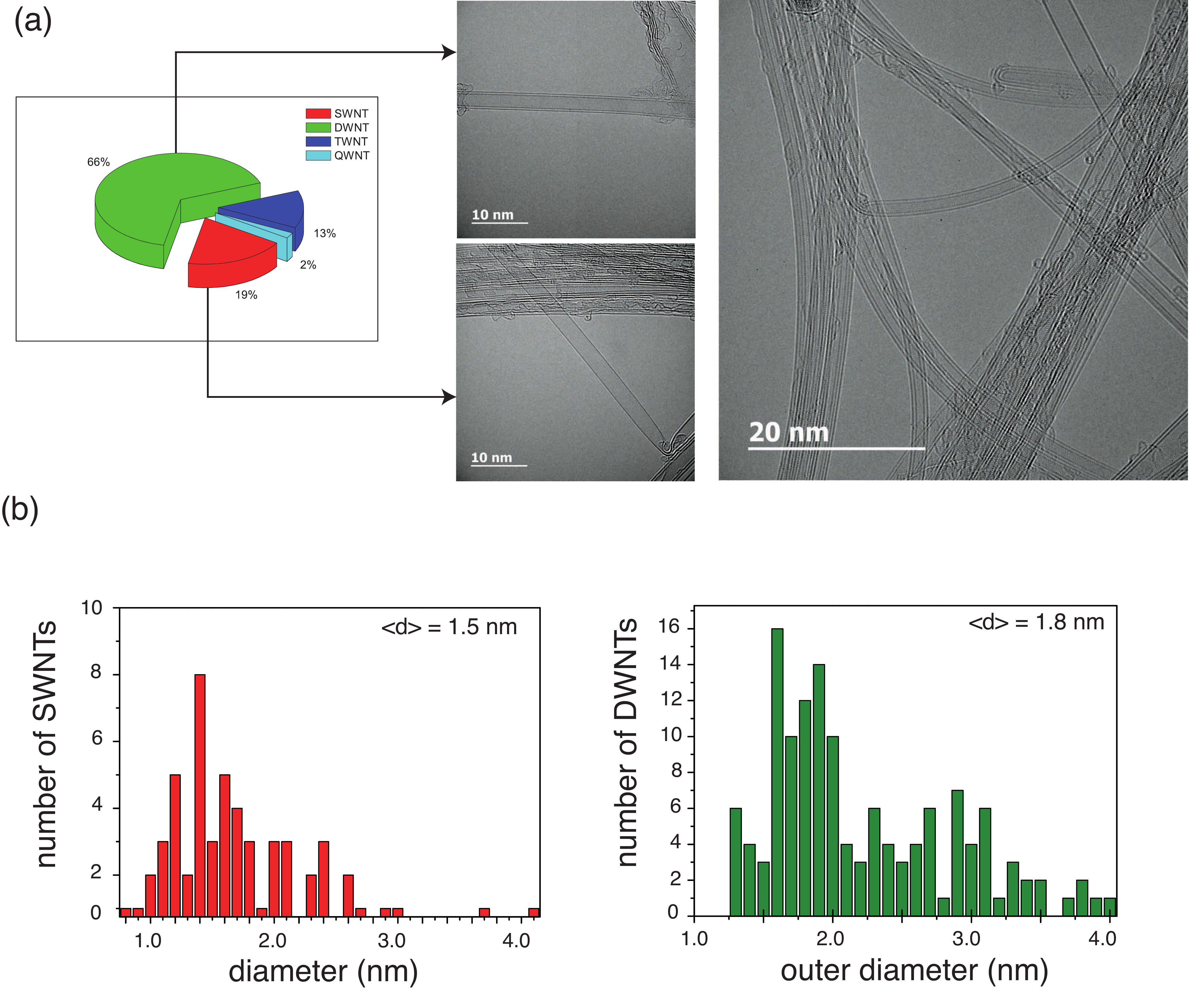}
\caption{(a) Population of different kinds of tubes (b) Histograms of all SWNTs and DWNTs. }
\label{figS1}
\end{center}
\end{figure}

\newpage

\textbf{Determination of tube diameters from TEM measurements}\\
With the help of projected potential and simulation images, the two diameters of the constituent tubes in a DWNT are measured. A simple measurement of the distance between dark lines pairs, corresponding to the tube walls, exhibits a systematic and substantial deviation from the correct $D$, even for a SWNT, due to the unavoidable Fresnel fringes arising at the edges of the nanotubes~\cite{Hashimoto2005, Loiseau2006}. In case of a DWNT, it is more complicated to measure the two diameters of the constituent tubes in a DWNT because of the more intense interference of Fresnel fringes between two adjacent tube walls. To get rid of such artefact, diameters have to be measured at the inversion point in the fringe profile. This can be done with the help of projected potential and simulation images and leads to precise $D$ with an error $\sim$0.05 nm (see Figure S2 of the supporting information). Using the approach presented in Figure \ref{figS2},  we can obtain an error $\sim$0.05 nm whatever the atomic resolution of the TEM we use.
\begin{figure}[htbp!]
\begin{center}
\includegraphics[width=0.3\linewidth]{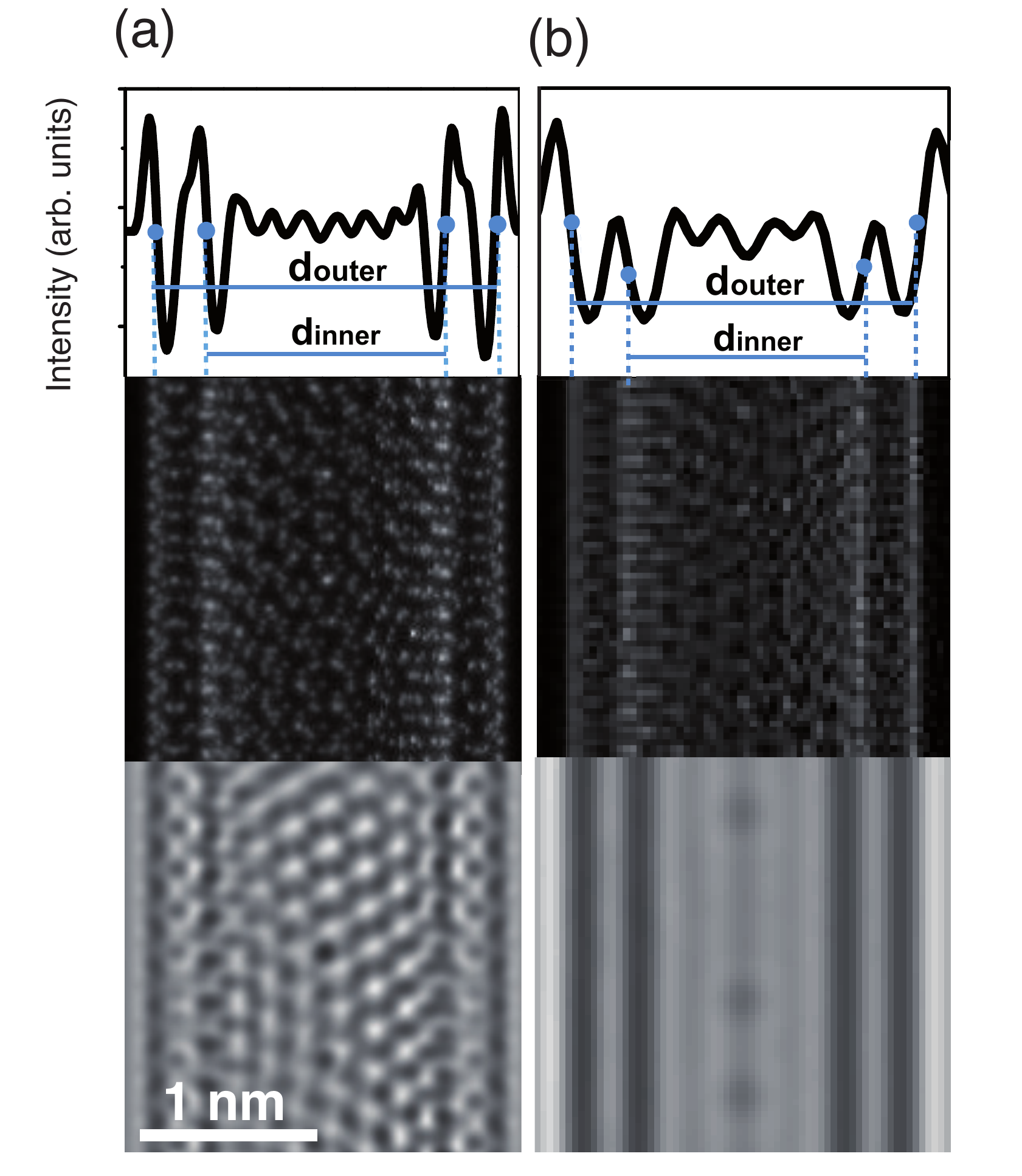}
\caption{ From bottom to top : simulated TEM image, projected potential image and profile of the equator line corresponding to the simulated TEM image. (a) New generation aberration-corrected microscope, the JEOL-ARM-200F with a spatial resolution of 50 $pm$ (b)  Conventional microscope, the Philips CM-20 with a spatial resolution of 270 $pm$.}
\label{figS2}
\end{center}
\end{figure}

\textbf{Reliability of the assignment}\\
Figure \ref{figS3} illustrates how Moir\'e patterns resulting from interferences of four walls in case of DWNTs are very sensitive to the $(n_{i},m_{i})$ and $(n_{o},m_{o})$ couples. A small change in the twist angle between inner and outer tube,  can significantly change the HR-TEM images. 

\begin{figure}[htbp!]
\begin{center}
\includegraphics[width=0.8\linewidth]{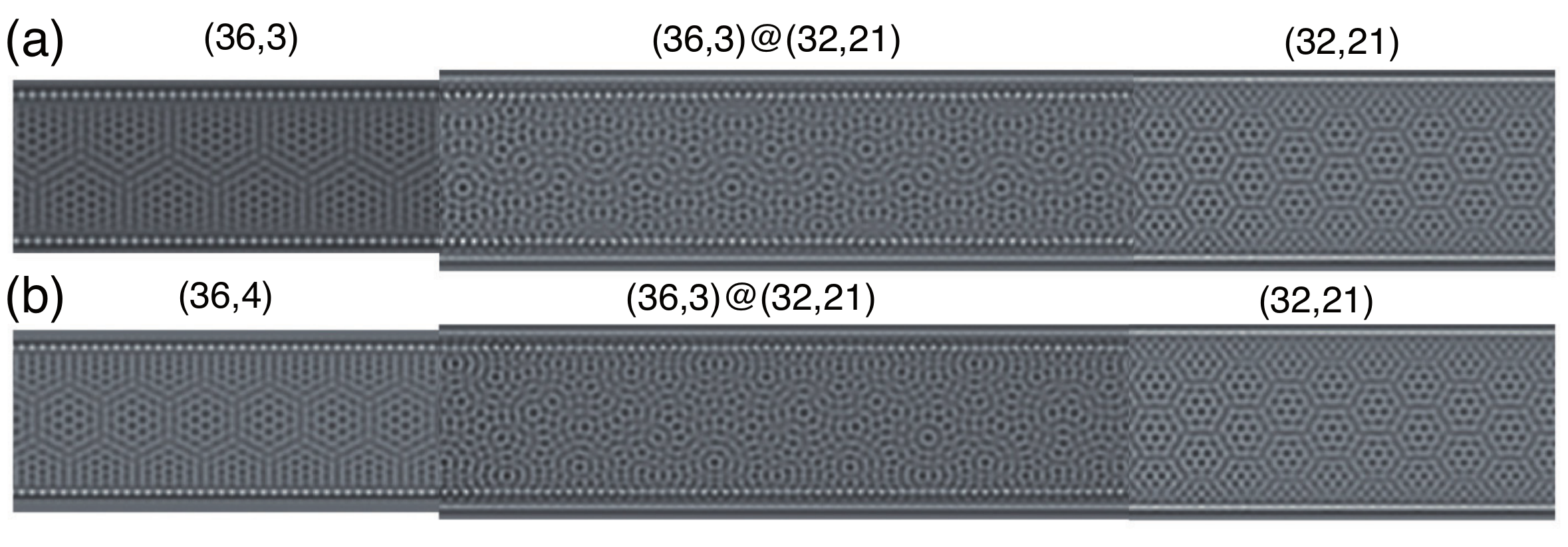}
\caption{Simulation images of (a) (36,3)@(32,21) DWCNT and (b) (36,4)@(32,21) DWCNT. }
\label{figS3}
\end{center}
\end{figure}

\textbf{Various examples of DWNT structure determination}\\
In Figure \ref{figS4} and \ref{figS5}, various examples are presented showing the procedure to determined unambiguously DWNT structures. 
 
\begin{figure}[htbp!]
\begin{center}
\includegraphics[width=0.75\linewidth]{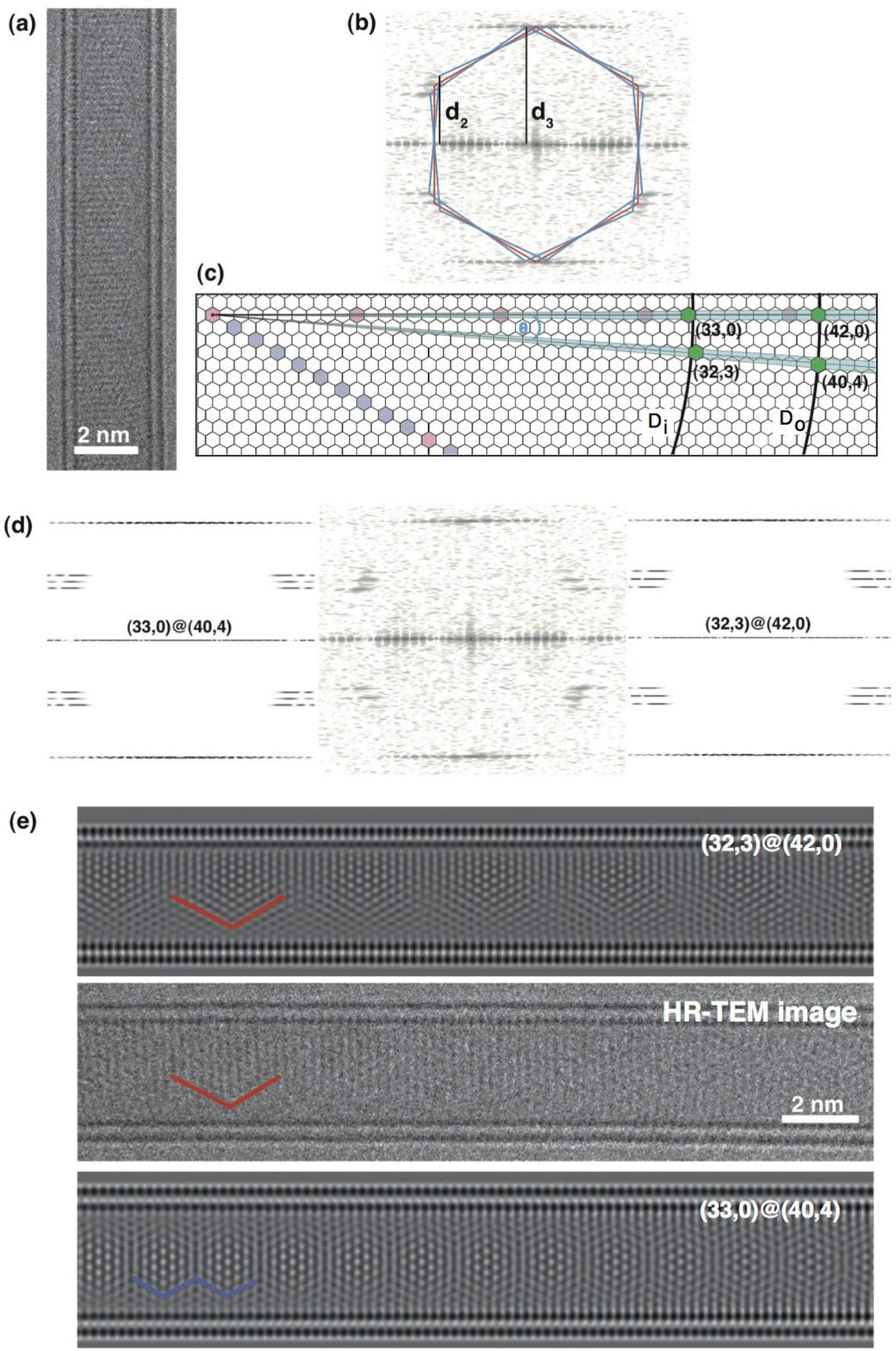}
\caption{Determination of a (32,3)@(42,0) DWNT ($\Delta\theta=4.43\degree$).(a) HR-TEM images of a DWNT and (b) its corresponding FFT. (c) Distribution of possible chiral indices after the analysis of the layer-lines. This lead to 2 configurations colored in green : (33,0)@(40,4) and (32,3)@(42,0). (d) Comparison of FFT from HR-TEM image and simulated results for previous solutions : no solution can be ruled out because differences are not significant. (d) Comparison of HR-TEM image and simulated images for deciding between the last configurations. Analysis of Moir\'e patterns, more precisely their sizes, enables to conclude that the investigated tube corresponds to (32,3)@(42,0) DWNT.}
\label{figS4}
\end{center}
\end{figure}
 
\begin{figure}[htbp!]
\begin{center}
\includegraphics[width=0.62\linewidth]{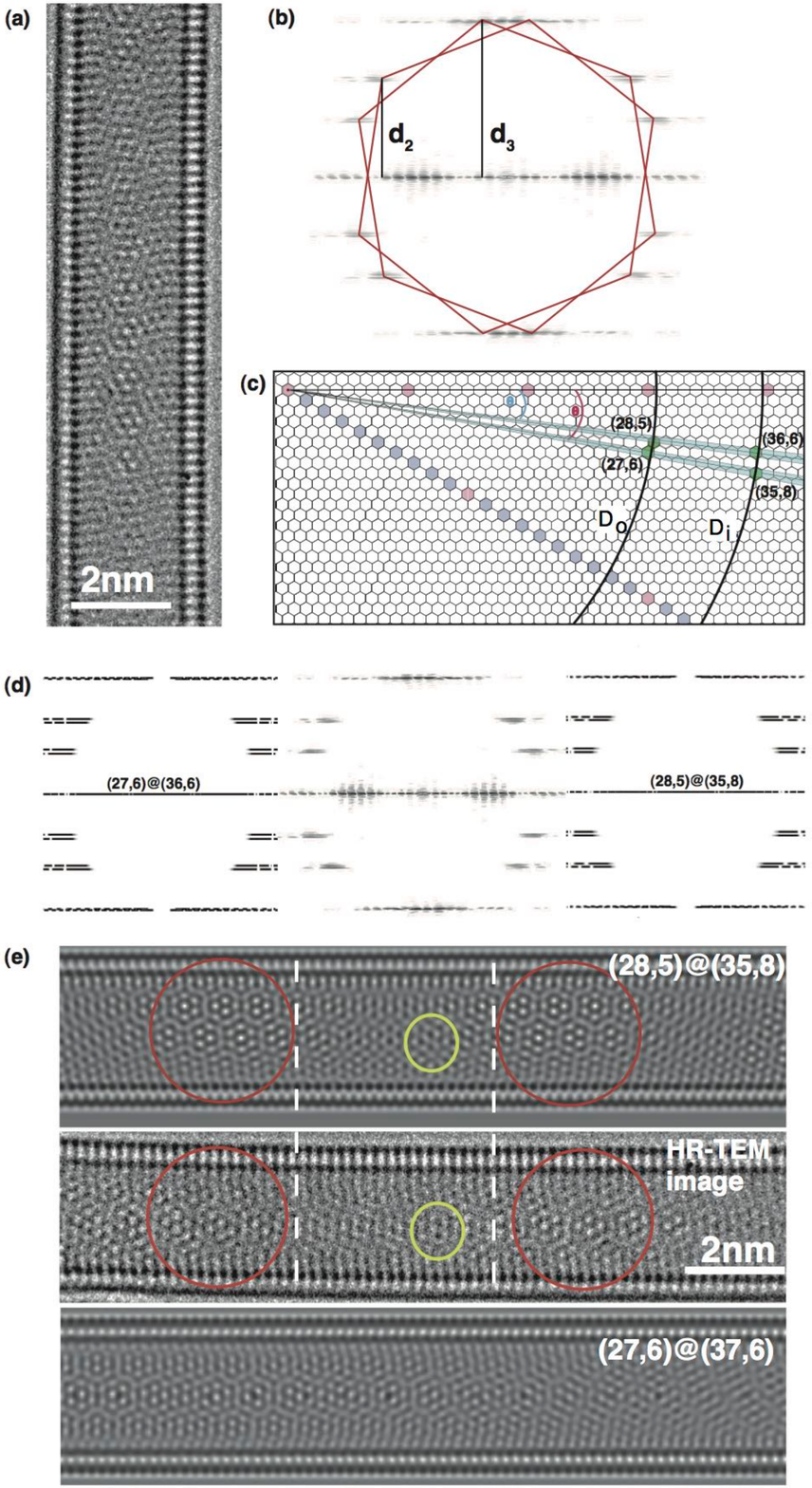}
\caption{ Determination of a (28,5)@(35,8) DWNT ($\Delta\theta\sim2.0\degree$). (a) HR-TEM images of a DWNT and (b) its corresponding Fourier transform. (c) Distribution of possible chiral indices after the analysis of the layer-lines. This lead to 2 configurations colored in green : (28,5)@(35,8) and (27,6)@(37,6). (d) Comparison of Fourier transform from HR-TEM image and simulated results for previous solutions : no solution can be ruled out because differences are not significant. (d) Comparison of HR-TEM image and simulated images for deciding between the last configurations. Analysis of Moir\'e patterns enables to conclude that the investigated tube corresponds to (28,5)@(35,8) DWCNT. }
\label{figS5}
\end{center}
\end{figure}

\newpage

\textbf{Distribution of structural parameters}\\
In Figure \ref{figS7}, we present the distribution of different structural parameters where it is clear that no correlation can be proposed.

\begin{figure}[htbp!]
\begin{center}
\includegraphics[width=0.70\linewidth]{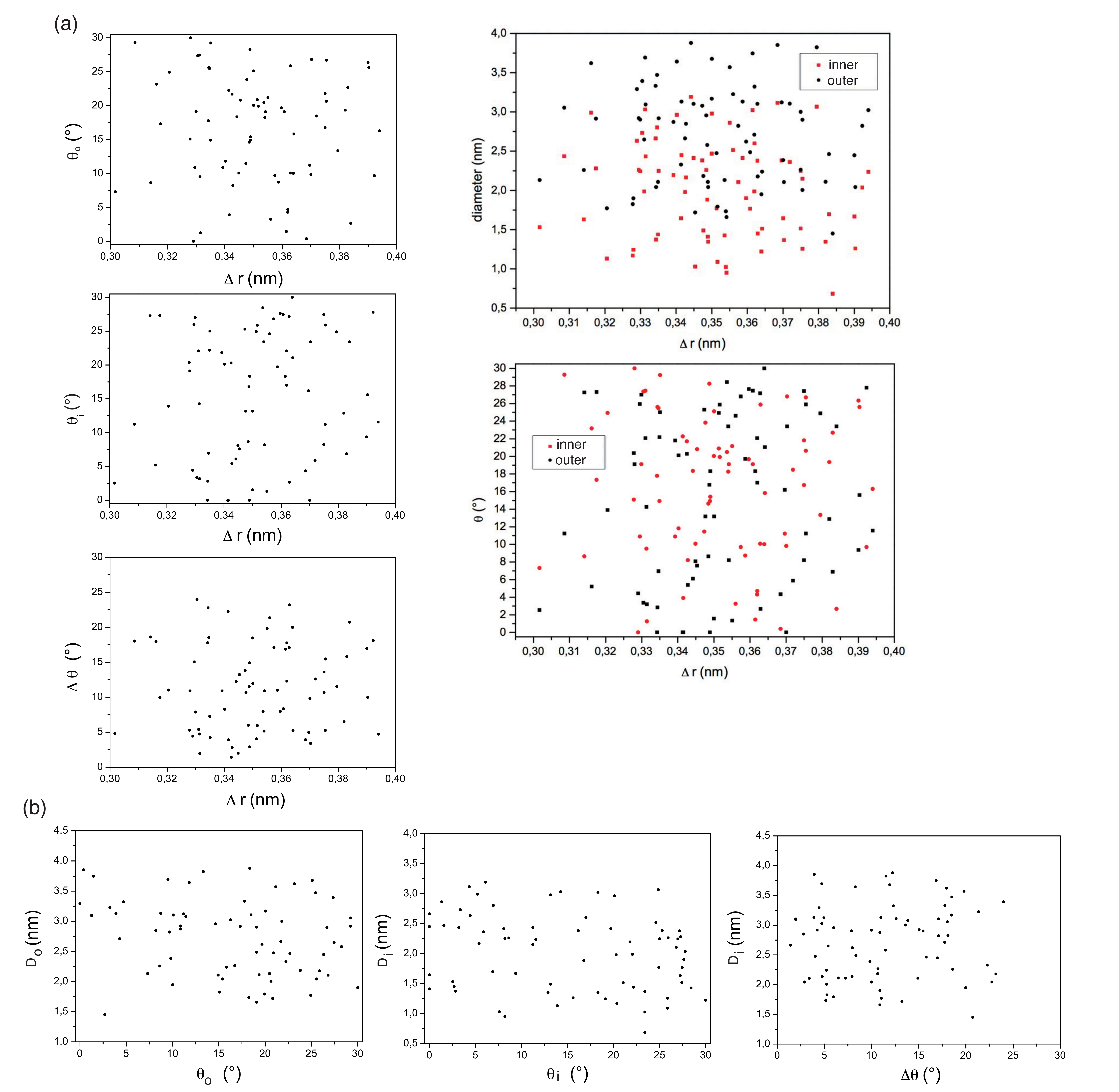}
\caption{Distribution of structural parameters ($\Delta r$, $\theta_{in}$, $\theta_{out}$ and $\Delta \theta$) of $\sim$ 70 DWNTs. (a) Helicities as a function of $\Delta r$. (b) Diameter as a function of $\Delta r$. (c) diameter as a function of helicities.}
\label{figS7}
\end{center}
\end{figure}

\newpage

\textbf{Analysis of data from the literature}\\
In Figure \ref{figS8}, we analyse the relationship between the helicities of outer tubes $\theta_{o}$ and $\Delta\theta$ from data found in the literature. Once again, forbidden configurations in grey areas ($\Delta\theta = 0\degree$ and $\Delta\theta > 25\degree$) have been found and a majority of DWNTs is depicted in a red square (both helicities are near armchair and $\Delta\theta < 15\degree$).

\begin{figure}[htbp!]
\begin{center}
\includegraphics[width=0.3\linewidth]{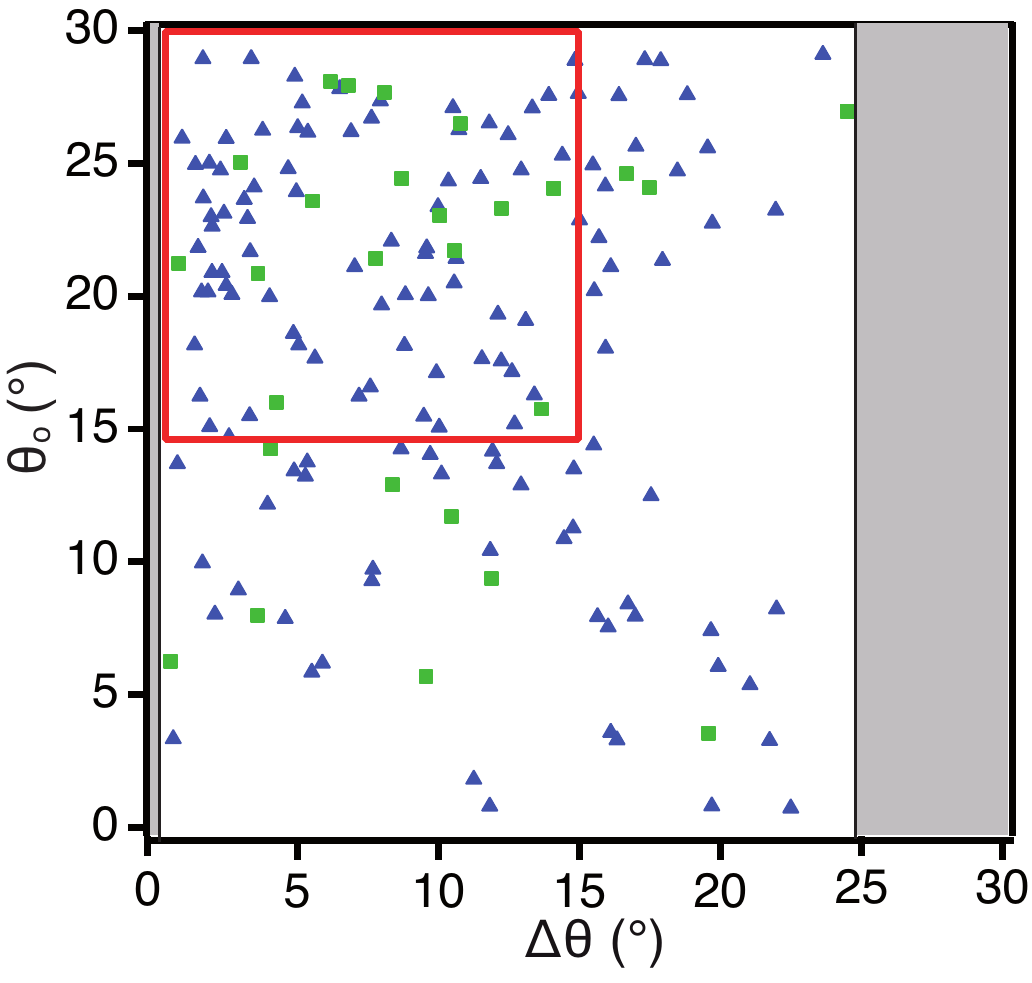}
\caption{Statistical analysis of DWNT helicities extracted from different experiments using different synthesis techniques : arc discharge (green squares from Ref.~\cite{Liu2014}) and CVD (blue triangles from Ref.~\cite{Hirahara2006}).}
\label{figS8}
\end{center}
\end{figure}

\newpage

\textbf{Analysis of observed and non observed configurations}\\
In Figure \ref{figS9}, analysis in term of local energies and C-C first neighbors intertube distances are presented for the observed (11,10)@(20,11) DWNT and non observed (8,8)@(13,13) DWNT.

\begin{figure}[htbp!]
\begin{center}
\includegraphics[width=0.70\linewidth]{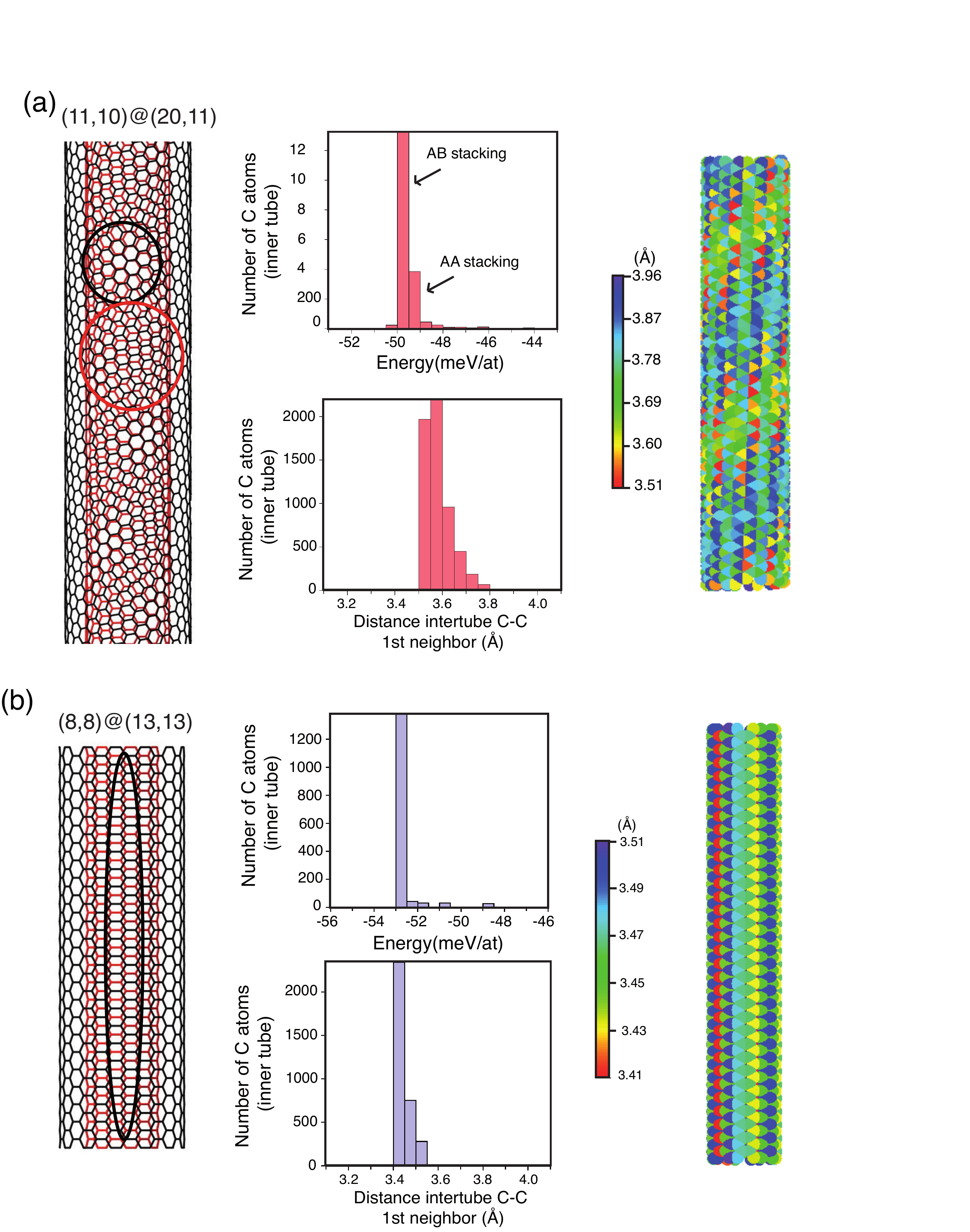}
\caption{Analysis of local energies in form of histograms plot (middle). Different stackings (AA or AB) are highlighted by circles (left) and analysis of the C-C first neighbors intertube distances in form of histogram plots (middle) and spatial distribution along the tube (right). (a) (11,10)@(20,11) DWNT ($\Delta\theta=20.48\degree$) and (b) (8,8)@(13,13) DWNT ($\Delta\theta=0\degree$).}
\label{figS9}
\end{center}
\end{figure}

\end{widetext}

\end{document}